# Lévy walk patterns in the foraging movements of spider monkeys (*Ateles geoffroyi*)

GABRIEL RAMOS-FERNÁNDEZ (✉), JOSÉ LUIS MATEOS, OCTAVIO MIRAMONTES, GERMINAL COCHO

*Departamento de Sistemas Complejos, Instituto de Física, Universidad Nacional Autónoma de México, Ciudad Universitaria, México 01000 D.F. México.*

HERNÁN LARRALDE

*Centro de Ciencias Físicas, Universidad Nacional Autónoma de México, Av. Universidad s/n, Col. Chamilpa, Cuernavaca 62210 Morelos, México.*

BÁRBARA AYALA-OROZCO

*Department of Environmental Studies, University of California, Santa Cruz, CA, 95064.*

Current address: Pronatura Península de Yucatán. Calle 17 #188A x 10, Col. García Ginerés. Mérida Yucatán 97070 México.

ramosfer@sas.upenn.edu

FAX: 52 999 925 3787

Telephone: 52 999 920 4647

## SUMMARY

Scale invariant patterns have been found in different biological systems, in many cases resembling what physicists have found in other nonbiological systems. Here we describe the foraging patterns of free-ranging spider monkeys (*Ateles geoffroyi*) in the forest of the Yucatán Peninsula, México and find that these patterns resemble what physicists know as Lévy walks. First, the length of a trajectory's constituent steps, or continuous moves in the same direction, is best described by a power-law distribution in which the frequency of ever larger steps decreases as a negative power function of their length. The rate of this decrease is very close to that predicted by a previous analytical Lévy walk model to be an optimal strategy to search for scarce resources distributed at random (Viswanathan et al 1999). Second, the frequency distribution of the duration of stops or waiting times also approximates a power-law function. Finally, the mean square displacement during the monkeys' first foraging trip increases more rapidly than would be expected from a random walk with constant step length, but within the range predicted for Lévy walks. In view of these results, we analyze the different exponents characterizing the trajectories described by females and males, and by monkeys on their own or when part of a subgroup. We discuss the origin of these patterns and their implications for the foraging ecology of spider monkeys.

# INTRODUCTION

Understanding how pattern and variability changes with the scale of analysis is one of the main goals of ecological research (Levin 1992). Ideally, one would like to extrapolate findings in one scale of analysis to another. This may be achieved by simplifying the problem at one scale, finding an expression that its main features without unnecessary detail and then using this expression to extrapolate to larger scales. Thus, the diffussion approximation to individual animal movements, where an individual is assumed to move at random as if it was a Brownian particle, has been successfully applied to predict large-scale features of the population such as its rate of spatial dispersal (Turchin 1998).

In some cases, universal scaling relationships which describe the same pattern at different scales have been found in biological systems (rev. in Gisiger 2001). Fractal geometry, where spatial patterns show statistically similar patterns over several orders of magnitude in the scale of observation, has been succesful in describing patterns of aggregation of resources (Ritchie and Olff 2000) and even in predicting the different relationships that should exist between body size and home range in different species (Haskell et al. 2002).

Here we describe the movement patterns of individual spider monkeys (*Ateles geoffroyi*) in the wild and find that they resemble what physicists have long recognized as Lévy walks. These walks show spatial scale invariance in the length of constituent steps and temporal scale invariance in the duration of intervals between steps. Similar movement patterns have already been found in the foraging flights of wandering albatrosses (*Diomedea exulans*; Viswanathan et al. 1996). The same authors developed an analytical model suggesting that Lévy walks are an optimal strategy for finding randomly distributed, scarce resources (Viswanathan et al. 1999).

In a Lévy walk (Schlesinger et al. 1993), the length of each successive step ($x$) varies according to a power-law function of the form:

$$N(x) \sim x^{-a} \qquad \text{where } 2 < a < 3$$

In other words, a Lévy walk has no intrinsic step length scale and thus steps of seemingly very long length may be observed. When taking the logarithm in both sides of the equation, the power-law relation appears as a straight line, implying that there is a constant proportion of steps of different lengths.

Lévy walks have been studied experimentally in many different fields in physics (Klafter et al. 1996). For instance, in fluid dynamics, the position of tracer particles is analyzed at regular intervals under different flowing media conditions (Weeks et al. 1995). In these experiments a particle can remain in the same location for some time before moving again. These waiting (or sticking) times also vary according to a power-law function of the form:

$$N(t) \sim t^{-b} \qquad \text{where } b \simeq 2$$

Because some very long steps may occur, in a Lévy walk the mean-squared displacement will not be a linear function of the time $t$ as it is in regular random walks (i.e. with constant step length). Rather, the mean-squared displacement $<R^2(t)>$ will vary as:

$$<R^2(t)> \sim t^{c} \quad \text{where } 1 < c < 2$$

In other words, the mean-squared displacement of a Lévy walker grows faster in time than that of a random walker with constant step lengths or one with a normal distribution of step lengths. In theoretical studies of Lévy walks, there is a precise prediction about the relationship that should exist between the three exponents presented above (Weeks et al. 1995):

$$\text{for } b > 2,\ c = 4 - a$$

$$\text{for } b < 2,\ c = 2 + b - a$$

Here we analyze the trajectories described by free-ranging spider monkeys in the rainforest of the northeastern Yucatán Peninsula, in México. We use the trajectories that 20 different individuals followed from dawn until dusk to analyze the frequency distribution of step lengths, the frequency distribution of the waiting time durations and the mean squared displacement at different times. We find that during a certain regime, spider monkey foraging trajectories are surprisingly similar to the Lévy walks studied by physicists.

We then use this finding as a tool to explore the possible variations in trajectories with regard to the grouping behavior of spider monkeys. This species forms temporary aggregations that vary in size and composition throughout the day, which have been suggested to occur in response to the variation in food availability (Klein and Klein 1977; Symington 1987, 1988). The trajectories described by lone individuals could then be different to those described by individuals in a subgroup, as they probably represent different strategies for finding or exploiting known food patches. In addition, male spider monkeys occupy larger home ranges, and travel farther per day and in larger subgroups than females (Symington 1987; Ramos-Fernández and Ayala-Orozco 2002) a difference that may reflect different space use strategies (Wrangham 2000). Therefore we analyzed the Lévy walk parameters separately in the trajectories of lone and grouped individuals as well as in those of females and males.

# METHODS

## Study site and animals

Data were collected in in the area of forest surrounding the Punta Laguna lake (2 km x 0.75 km), in the Yucatán Peninsula, México ($20^{o}$38' N, $87^{o}$38' W, 14 m elevation). This region is characterized by seasonally dry tropical climate, with mean annual temperature of about $25^{o}$C and mean annual rainfall around 1500 mm, 70% of which is concentrated between May and October. The main forest fragment near the lake consists or 60 hectares of medium semi-evergreen forest. This is in turn surrounded by secondary successional forest about 30-40 years old in an area of 5000 hectares recently declared as a protected area, the Otoch Ma'ax Yetel Koh Sanctuary. Spider monkeys use both of these vegetation types, although they spend more than 70% of their daily time and every night in the medium forest (Ramos-Fernández and Ayala-Orozco 2002). Trails were cut throughout the fragment of medium forest and through part of the successional forest. In these trails, trees and other landmarks were used to make accurate maps of this area. Visibility conditions were very good as monkeys used the canopy at heights from 5 to 25 m. More details about the study site can be found in Ramos-Fernández and Ayala-Orozco (2002).

Two study groups, occupying 0.95 and 1.66 $km^2$ of forest, respectively, have been studied continuously since January, 1997. One group (5 female and 3 male adults as of 1999) was habituated to human presence before the study began and the other (15 female and 6 male adults as of 1999) was habituated during 1997. All monkeys were identified by facial marks and other unique features. Adults were defined by their size and darker faces, and in the case of males, for their fully descended testes. All monkeys could be reliably identified by the end of 1997.

## Data collection

Data reported here were collected between September through December 1999. On 20 days during that period, a different known adult was chosen as the focal subject and followed by at least two observers from dusk until dawn, taking an instantaneous sample of its location, activity and subgroup size and composition every 5 minutes. The location was estimated visually by two observers to the nearest 5 m with respect to landmark trees or paths. In all cases a landmark was closer than 50 m from the position of the monkey.

## Data analysis

Trajectories were analyzed according to the methods outlined by Turchin (1998). The trajectory of each focal monkey consisted of a sequence of paired coordinates, one pair for each 5 minute interval where an instantaneous sample had been recorded. A step was defined as an interval in which any or both of the coordinates in two consecutive samples differed. The length of each step was the linear distance separating the position at two consecutive samples. In some cases, observers lost sight of the focal animal for a number of sample intervals. Steps were not calculated for those intervals but only for those in which the position was known for two consecutive five-minute samples. The frequency distribution of step lengths was analyzed using a bin size of 10 m. The log-log plot of this distribution was used to calculate the relationship that produced the best fit using a least-square method.

Wating times were calculated from the number of samples in which the focal animal did not change position. The frequency distribution of waiting times was analyzed using a bin size of one interval (5 minutes). The log-log transform of this distribution was used to calculate the relationship producing the best fit using a least-square method.

Squared displacements were calculated by measuring, for each individual separately, the length of a line joining its location in the first sample of the day and its location at different times thereafter. The mean-squared displacement was obtained by averaging all squared displacements accross all individuals for a given time of day. From this mean-squared displacement a maximum was obtained around 1030 hours, which then decreased as monkeys consistently began approaching their starting point, in some cases returning to it at noon and in most cases returning to it at dusk (see Figure 4). The mean-squared displacement from 0630 to 1030 was then taken to be a period in which most of the individuals moved away from their starting point. For this period only, a log-log plot was produced and a line adjusted by the least-square method.

Turning angles were calculated by substracting the absolute angle (with respect to the east-west axis and in a counterclockwise direction) of each step from the absolute angle of the previous step. A frequency distribution of turning angles was thus produced, using a bin size of 10 degrees.

The regression slopes for the Lévy walk distribution of trajectories of lone vs. grouped individuals and females vs. males were compared using an F test for regression slopes (Sokal and Rohlf 1994).

## RESULTS

The observed daily trajectories of spider monkeys are made up of steps of variable length. Three examples of these trajectories are shown in Figure 1(a-c). One important property of Lévy walks is that they show self-similarity accross different spatial scales. Indeed, when we close in on a region of one of these trajectories, a qualitatively similar pattern as in the large scale appears (Figure 1d).

**Figure 1**

The frequency distribution of step lengths of all trajectories together shows a wide range of variation (Figure 2a). The data can be fitted better to a power-law function (Figure 2b; $r^2$ = 0.89) than to an exponential one ($r^2$ = 0.79). The value of the exponent in the negative power-law equation fitted to the data is $a$ = 2.18.

**Figure 2**

The frequency distribution of waiting times of all trajectories together also shows a wide degree of varation. Before traveling again, spider monkeys may stop for as little as 10 minutes or for as much as 2 hours (Figure 3). The log-log plot of the data is fit well by a negative power-law function with an exponent $b$ = 1.7 ($r^2$ = 0.86).

**Figure 3**

The squared displacement in many of the individuals analyzed shows a common pattern: monkeys tend to get away from their sleeping site for a few hours in the morning, before coming back to the origin at noon or staying in the vicinity, normally coming back to the same sleeping site shortly before sunset. Some individuals, especially males, did not return to the same sleeping site at the end of the day. Close inspection of the mean-squared displacement of all individuals shows a maximum at around 1030 hours (Figure 4a). A log-log plot of the data for the period between 0630 and 1030 only, when all individuals were getting away from the origin, adjusts to a line with a slope of 1.7 ($r^2$ = 0.87), as would be expected in a Lévy walk (Figure 4b).

**Figure 4**

The fact that spider monkeys travel back to their sleeping sites, as well as the fact that they sometimes use the same routes for going away from and returning to their sleeping site (for example, see Figure 1b), implies that some persistence should exist in the direction of consecutive steps. Figure 5 shows the distribution of turning angles between successive steps. This distribution is far from being uniform, being centered around zero. However, at a large enough time scale, this persistence is not expected to affect the scaling relations between the mean-squared displacement and time.

**Figure 5**

Spider monkeys change subgroup several times during a day, so the trajectory described by one individual in one complete day includes some steps traveled on its own and others when in a subgroup. When analyzing the distribution of lengths for these steps separately, the value of the exponent is different: $a_s = 1.50$ ($r^2 = 0.80$) for solitary individuals and $a_g = 2.12$ ($r^2 = 0.89$) for individuals in groups (Figure 6). An F test comparing the two regression slopes shows a significant difference ($F = 5.72$, $P < 0.05$). Considering the fact that the number of samples in each category are different, monkeys travel a higher proportion of long trajectories when on their own than when part of a subgroup (Figure 6).

**Figure 6**

We also analyzed the distribution of travel lengths separately for females and males. Figure 7 shows that both are described by power laws with exponents $a_f = 2.11$ for females and $a_m = 1.47$ for males. It is males who travel a higher proportion of long trajectories than females.

## DISCUSSION

We have presented evidence showing that the daily movements of spider monkeys resemble what physicists know as Lévy walks, i.e. random walks with power-law scaling in the length of their constituent steps. Waiting times, or the duration of intervals without movement, also show power-law scaling. As expected from theoretical studies of Lévy walks, the mean-squared displacement increases faster in time than in other random walks such as a Brownian random walk (Schlesinger et al. 1993).

In support of this interpretation, there is an agreement between all three exponents, as would be predicted from theoretical studies of Lévy walks (Weeks et al. 1995):

$$\text{for } b < 2, \; c = 2 + b - a$$

which, substituting the observed values of each exponent, yields the following prediction for the value of c:

$$\text{for } b = 1.7 < 2$$

$$c = 2 + 1.7 - 2.18 = 1.52$$

A value that is close to the observed value of 1.7. In other words, there is an agreement between different properties of the observed trajectories during this time lapse, such that they can accurately be described as Lévy walks.

Lévy walks differ in fundamental ways from other types of random walks, such as Brownian random walks, which have been used in most models of animal foraging (Turchin 1998). As mentioned before, the mean-squared displacement after a time *t* is larger for Lévy walks than for other random walks (Weeks et al. 1995). In the period between 0630 and 1030, when spider monkeys normally spend two or three hours foraging away from their sleeping sites (Ramos-Fernández, personal observations), the mean-squared displacement increases faster than in a Brownian random walk.

However, there are certain regularities in the trajectories analyzed here that would not allow us to label them as “pure” Lévy walks. After 1030 hours spider monkeys tend to stop traveling away from their sleeping sites and on most ocassions return to them before sunset. Also, the shape of their home range, which circles a lake (e.g. see Figure 1c), implies that sometimes a monkey will use the same route to go in one direction and return in the opposite direction. The idea that monkeys are actually following a route of some sort is suggested by the fact that two consecutive steps tend to be given in a similar direction, as shown in the distribution of turning directions, which is centered around zero.

Interpretation of these results hinges on the crucial question of the cues that spider monkeys use for finding food sources. Feeding on the fruit from 50 to 150 species of trees (van Roosmalen and Klein 1987; Ramos-Fernández and Ayala-Orozco 2002), each with its own phenological pattern, spider monkeys clearly face the problem of exploiting an extremely variable resource. Lévy walks could represent a foraging search strategy, in which the movements of spider monkeys are guided by the odour and visual cues they detect from the existing ripe fruit in fruiting trees. Alternatively, the patterns reported here could be the result of the knowledge based on memory that spider monkeys could have about the location of fruiting trees.

Little is known about what animals in general know about the location of food sources (rev. in Boinski and Garber, 2000). Experimental evidence suggests that wild capuchin monkeys (*Cebus apella*) visit the closest food sources more often than would be expected on the basis of "random search" null models (Janson 1998). Also, they appear to use straighter lines than would be expected if they were only searching with no memory of where they had found food in the past. If spider monkeys have a similar kind of spatial knowledge about the past location of food, then the patterns that we report here may not represent "random" searches at all, but the result of more directed travel between known food sources.

However, the null models developed by Janson (1998) as an expectation of travel with no spatial knowledge assume either a constant step length and arbitrary changes in direction or a constant travel direction until finding a food source. In the results we report here, spider monkeys are in fact including some very long steps in their foraging trajectories. If this is a common pattern in tree-dwelling monkeys, then a more appropriate null model for foarging with no memory should include visits to farther, unknown food sites. This, in Janson's (1998) study, would make it more difficult to distinguish between observed and expected visiting frequencies.

It is possible that spider monkeys travel in Lévy walks in order to exploit these unknown sources of food more efficiently. Viswanathan et al. (1999) have shown that a Lévy walker visiting a number of randomly distributed foraging sites will visit more new sites and revisit less previously visited sites than a Brownian random walker traveling the same total distance. This is because the long steps in a Lévy walk quickly take the forager to more distant sites, making it less likely that it will walk on its own steps again. In this study, the value for the exponent $a$ in the distribution of step lengths is close to the optimum of $a = 2$ predicted by Viswanathan et al (1999) for foragers searching for randomly and sparsely distributed food sources.

Lévy walk foraging patterns could also be the result of the distribution of fruiting trees themselves. In an extensive study on several species of tropical trees in different sites around the world, Condit et al. (2000) found highly aggregated distribution patterns for most of the species, showing graphs of density of neighbors as a function of distance that look strikingly similar to those described by power-law functions. Also, Solé and Manrubia (1995) report self similar distributions of tree gaps (and thereby sites available for establishment of many new individual trees) in a tropical forest in Panamá. It is possible that at any given time, the fruiting trees where spider monkeys can feed are distributed in a scale-invariant, fractal manner. Thus, the long steps in the Lévy walks could be those given between patches of a given fruiting tree, while shorter steps would be given while foraging within a patch. Clearly, more information on the spatial distribution of the spider monkeys' food resources is required to clarify this issue.

Because spider monkeys are important seed dispersers for several tree species, in reality there could exist a bidirectional relationship between their foraging patterns and the distribution of trees. By foraging and dispersing seeds in such a pattern, spider monkeys might favor, in the long run, a self similar distribution of the very same trees on which they feed. A similar positive feedback between foraging behavior and the distribution of resources was found by Seabloom and Reichman (2001) in a simulation model of gopher – grass interaction. By restricting their movements to their defended territories, gophers disturbed areas where more grass could grow on the following season, therefore increasing habitat patchiness, which increased the foraging efficiency of gophers in the long run. Westcott and Graham (2000) report a long-tailed distribution in the distance traveled by a tropical flycatcher (*Myonectes oleaginus*) continuously after feeding. Based on the times of digestion for the seeds of different species, the authors derive the expected shape of seed shadows for different species, which also look very similar to power laws.

Another consequence of foraging in a Lévy walk pattern is that previously visited food sources may be revisited after long periods of time, favoring the ripening of more fruit before the next visit. Such “harvesting” of resources was suggested by Janson (2000) to be one of the reasons why animals foraging on scarce patches of food would not necessarily visit the closest one in the most optimal route. Perhaps the temporal dimension of the Lévy walk patterns reported here is as important as the spatial one.

We have found that the Lévy walks of females and males are different. Males have a larger proportion of long trajectories than females. This is consistent with the fact that male spider monkeys range over wider areas and travel further per day than females (Symington, 1987; Ramos-Fernández and Ayala-Orozco 2002). While males range over the whole territory of their group, females seem to restrict most of their time to a portion of it (Ramos-Fernández, personal observation). At the boundaries of their group’s territory, males form coalitions that engage in very aggressive encounters with neighboring males (Symington 1987; Ramos-Fernández, personal observations). In view of these results, the fact that the Lévy walks of males contain more long steps than those of females is consistent with their different space use strategies.

Finally, we have found a different value of the power-law exponent for the length distribution of steps given by lone monkeys as for those given by monkeys when part of a subgroup. In particular, monkeys on their own seem to travel a higher proportion of long steps compared to short ones. One of the cited benefits of group foraging has been an improvement on the likelihood that the group would find food patches which an individual on its own would not find (Krebs and Davies 1993). If spider monkeys do not have knowldege of the location of food sources, they could still find more fruting trees when traveling in a subgroup than when alone. The same argument applies if spider monkeys have knowledge about the location of fruiting trees: if they can share information on the known location of resources (Boinski and Garber 2000) it would seem logical that a subgroup would find more fruiting trees than a single individual, therefore decreasing the proportion of long steps.

The origin of the grouping pattern of spider monkeys has led to some discussion (Symington 1990; Wrangham 2000, in chimpanzees and spider monkeys). An intriguing possibility is that, in a hypothetical ancestor with stable grouping patterns, a Lévy walk foraging pattern increased the chances of separating from the rest of the group when foraging on scarce resources. In a group of *n* Lévy walkers, the probability that individuals will remain near the origin (or cross each other's path if they all start together at the same spot) is greatly reduced compared to a group of *n* Brownian walkers (Larralde et al 1992). Favored by the decrease in feeding competition offered by Lévy walks, social behaviors that maintained group membership even in the absence of visual contact would then develop, leading to the fission-fusion grouping pattern that we see today.

## ACKNOWLEDGEMENTS

The authors are truly grateful to Eulogio and Macedonio Canul, who provided invaluable assistance in the field by locating the study subgroups and collecting data. Laura Vick and David Taub initiated the study of spider monkeys in Punta Laguna. Fieldwork was financed by a graduate scholarship from the National Council for Science and Technology (CONACYT, Mexico) and by grants from the Wildlife Conservation Society, the National Commission for the Knowledge and Use of Biodiversity (CONABIO, Mexico), the Mexican Fund for the Conservation of Nature (FMCN) and the Turner Foundation. Data analysis and manuscript preparation was financed by CONACYT project 32453-E and DGAPA project IN-111000, as well as a visiting scholarship (Cátedra Tomás Brody) from the Complex Systems Department at the Physics Institute, National Autonomous University of México. These observations were performed in compliance with the Mexican Environment Protection Law (LGEEPA).

# FIGURE LEGENDS

**Figure 1** Daily trajectories of spider monkeys. (a) and (b) adult females. (c) adult male, with the section of the trajectory within the lower-left square amplified in (d). Note that some individuals, like the adult female in (b), returned to sleep close to where they started their daily travel.

**Figure 2** Distribution of the number of 5-minute intervals N(x) during which spider monkeys traveled a distance of x meters. A total of 841 five-minute intervals from 20 adult individuals are included. The insert (b) shows the log-log plot of the same data. A power-law relationship fits the data with $r^2 = 0.89$. The estimated value of the exponent is -2.18.

**Figure 3** Distribution of waiting times in the trajectories of spider monkeys. The figure shows the log-log plot of the number of intervals $N(t)$ with duration $t$. The relationship is fit by a power-law function with an estimated value of the exponent of -1.7 ($r^2 = 0.86$).

**Figure 4** Squared displacement in the trajectories of spider monkeys. (a) the mean squared displacement across all individual trajectories at different times of day. Note that there is a maximum at 1030 hours. (b) log-log plot of the mean squared displacement observed from 0630 to 1030 hours. The relationship is well fit by a power-law function with an estimated value for the exponent of 1.7 ($r^2 = 0.87$).

**Figure 5** Circular distribution plot of the turning angle between consecutive steps. Shown is the number of times that two consecutive steps in the trajectories differed by the degrees shown. Note that the distribution peaks around zero.

**Figure 6** Distance traveled by spider monkeys when alone and in groups. Shown is a log-log plot of the number N(x) of five-minute intervals in which a lone adult spider monkey traveled a distance x. For monkeys traveling alone, a total of 156 five-minute intervals are included. A power-law relationship fits the data with $r^2 = 0.8$. The estimated value of the exponent is -1.50. For monkeys traveling in a group, a total of 685 five-minute intervals are included. A power-law relationship fits the data with $r^2 = 0.89$. The estimated value of the exponent is -2.12.

**Figure 7** Distance traveled by female and male spider monkeys. Shown is a log-log plot of the number N(x) of continuous trajectories traveled by adult spider monkeys without stopping, for a distance of x meters. For 14 females, a total of 604 five-minute intervals were analyzed. A power-law relationship fits the data with $r^2 = 0.9$. The estimated value of the exponent is -2.11. For 7 males, a total of 237 five-minute intervals were analyzed. A power-law relationship fits the data with $r^2 = 0.93$. The estimated value of the exponent is -1.47.

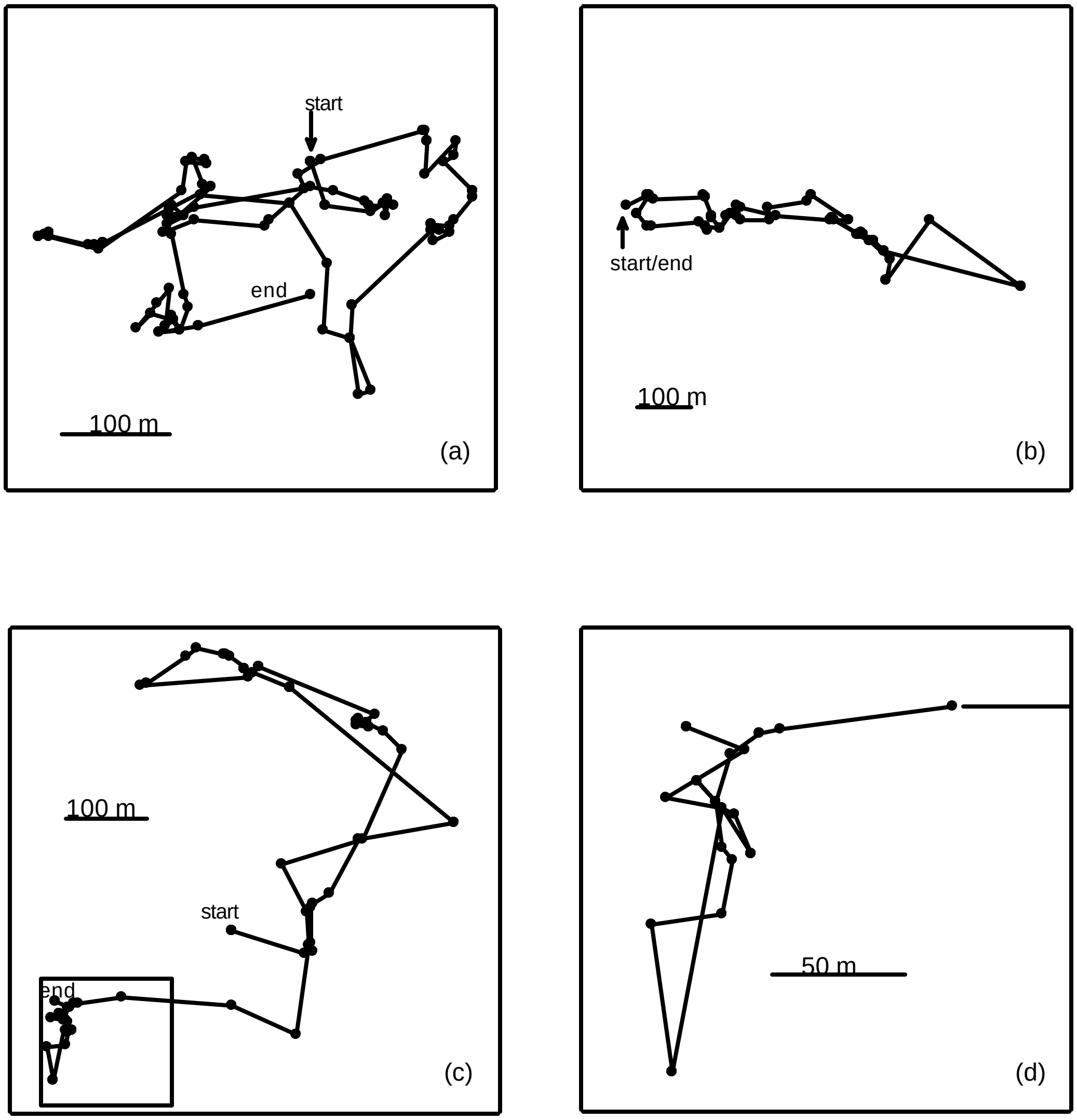

Figure 1

Figure 2

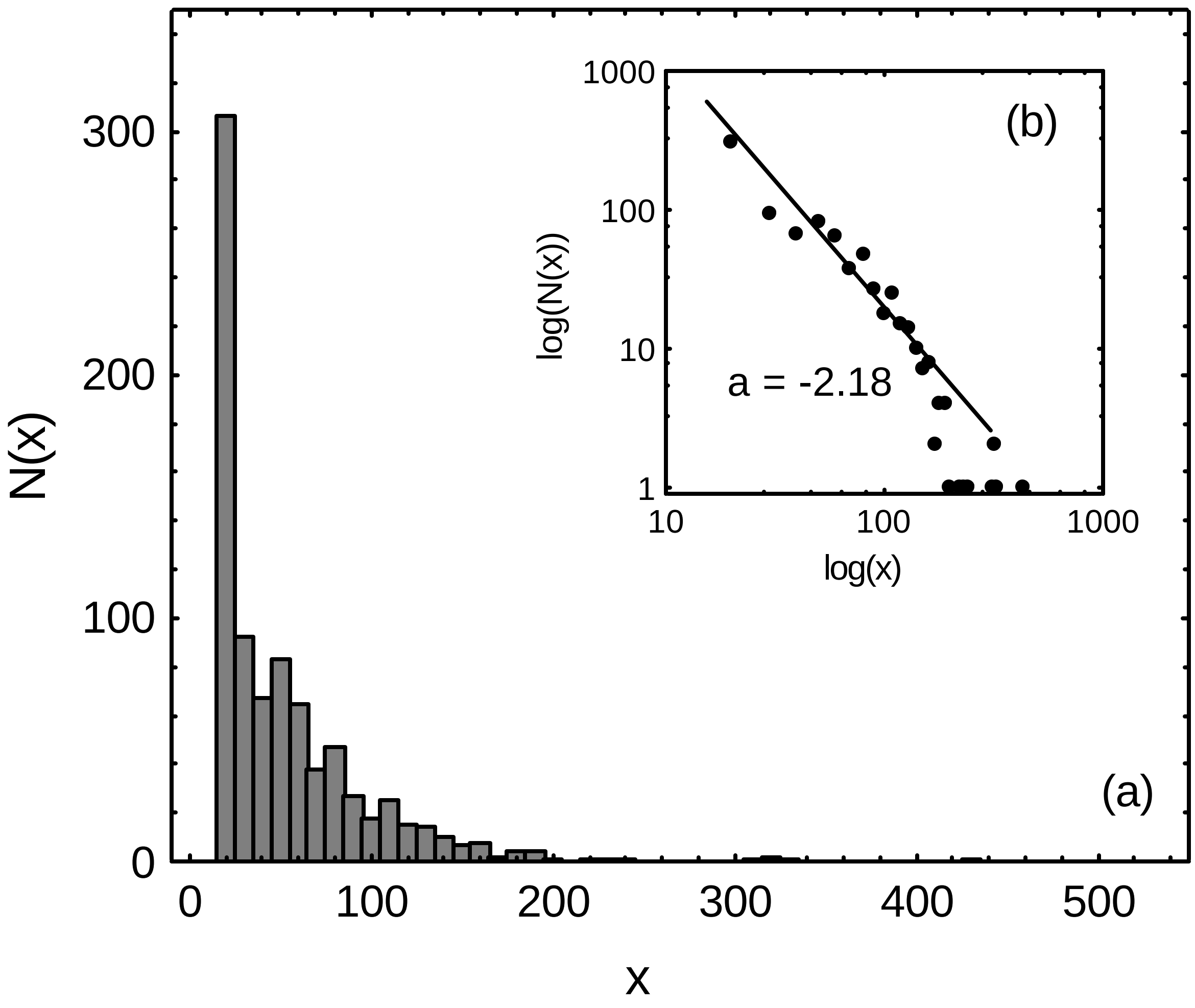

Figure 3

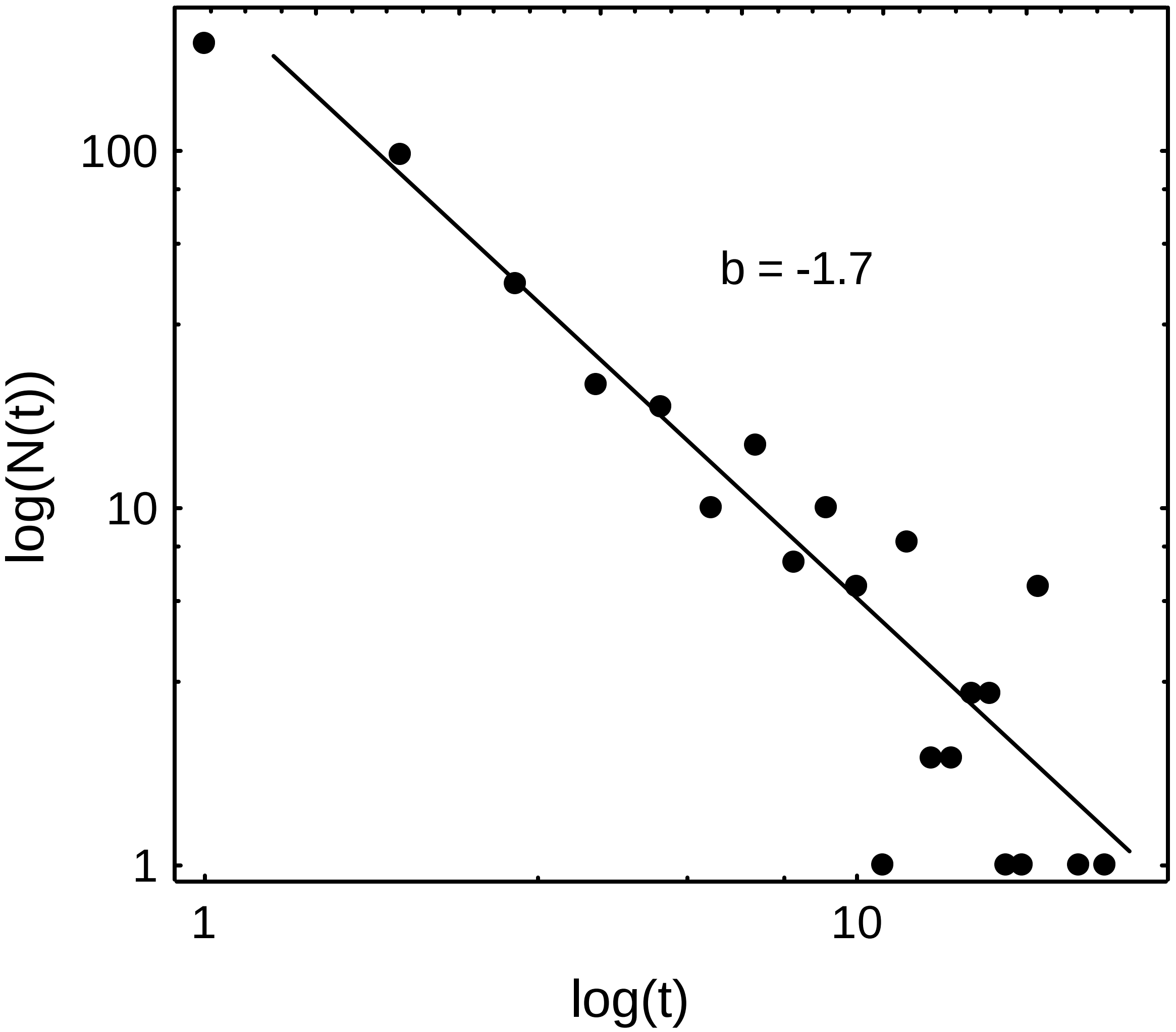

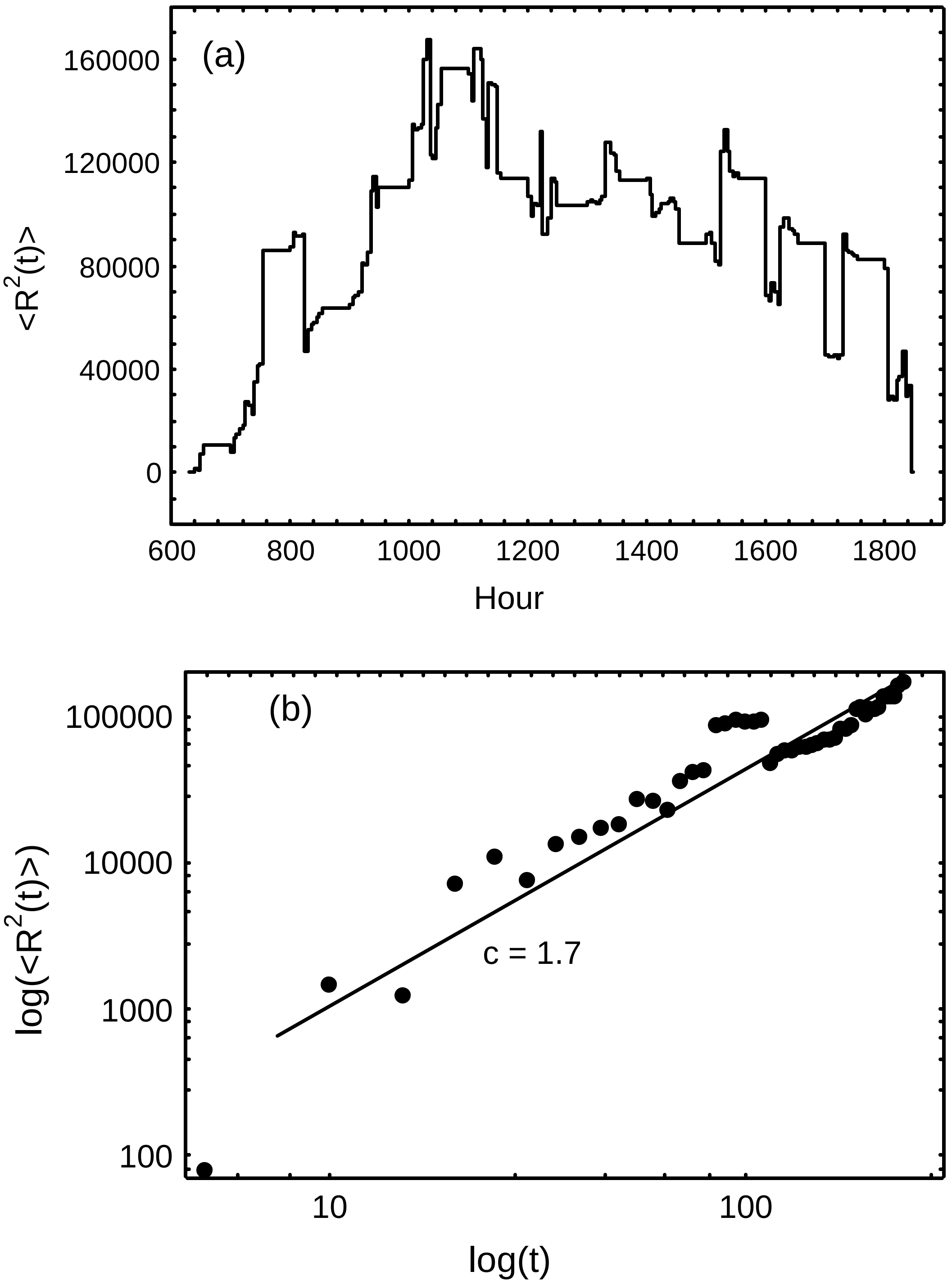

Figure 4

Figure 5

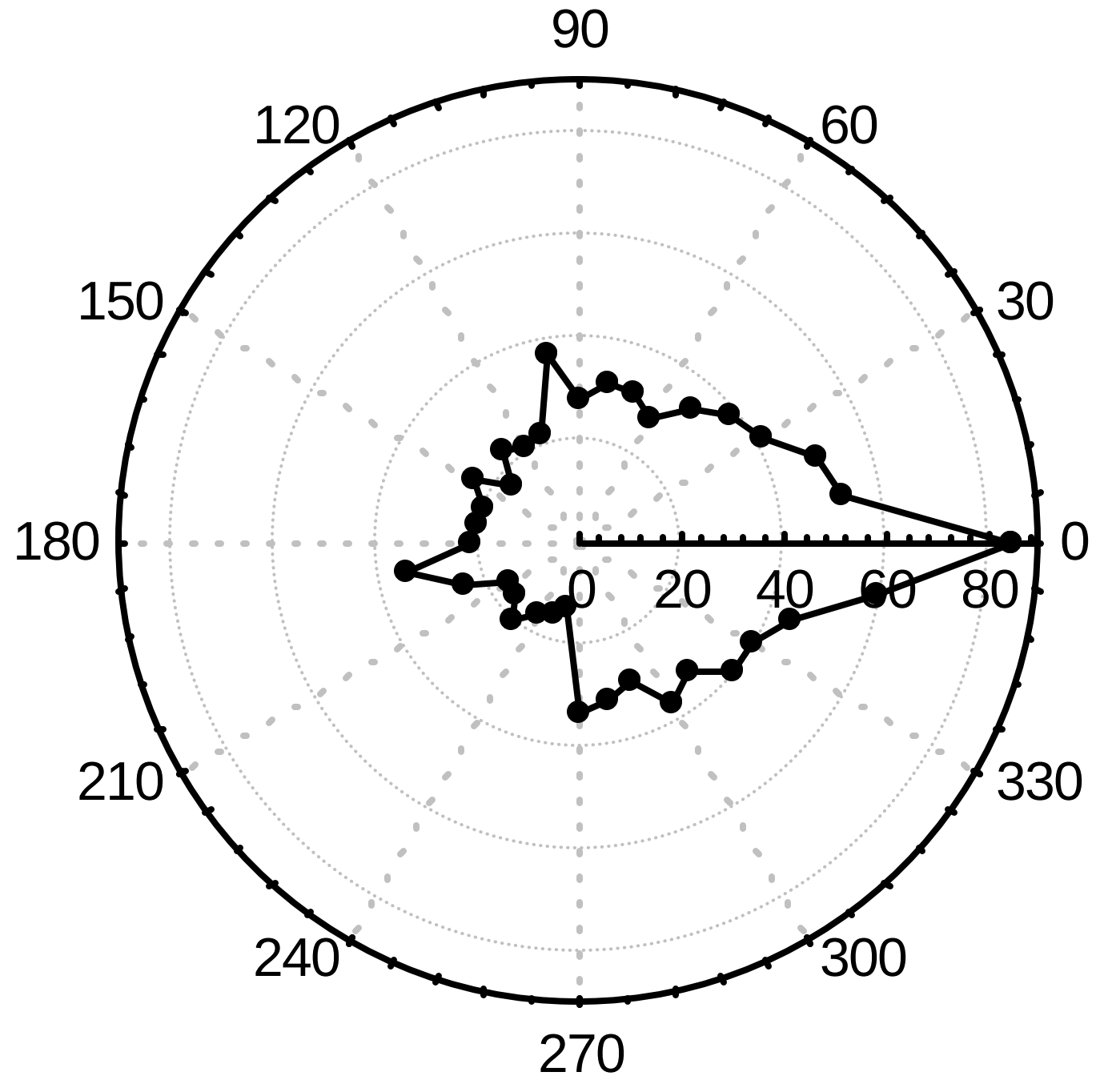

Figure 6

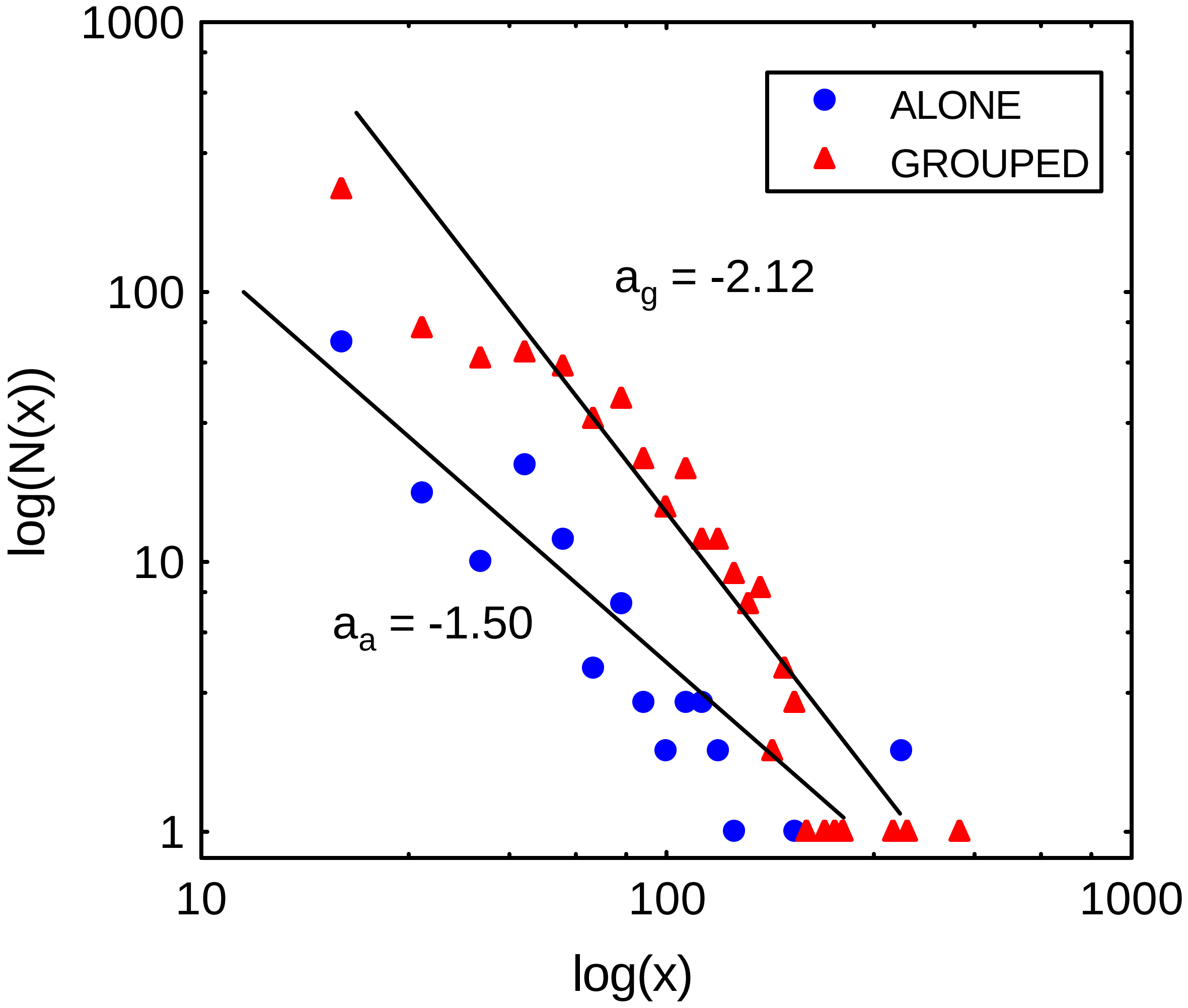

Figure 7

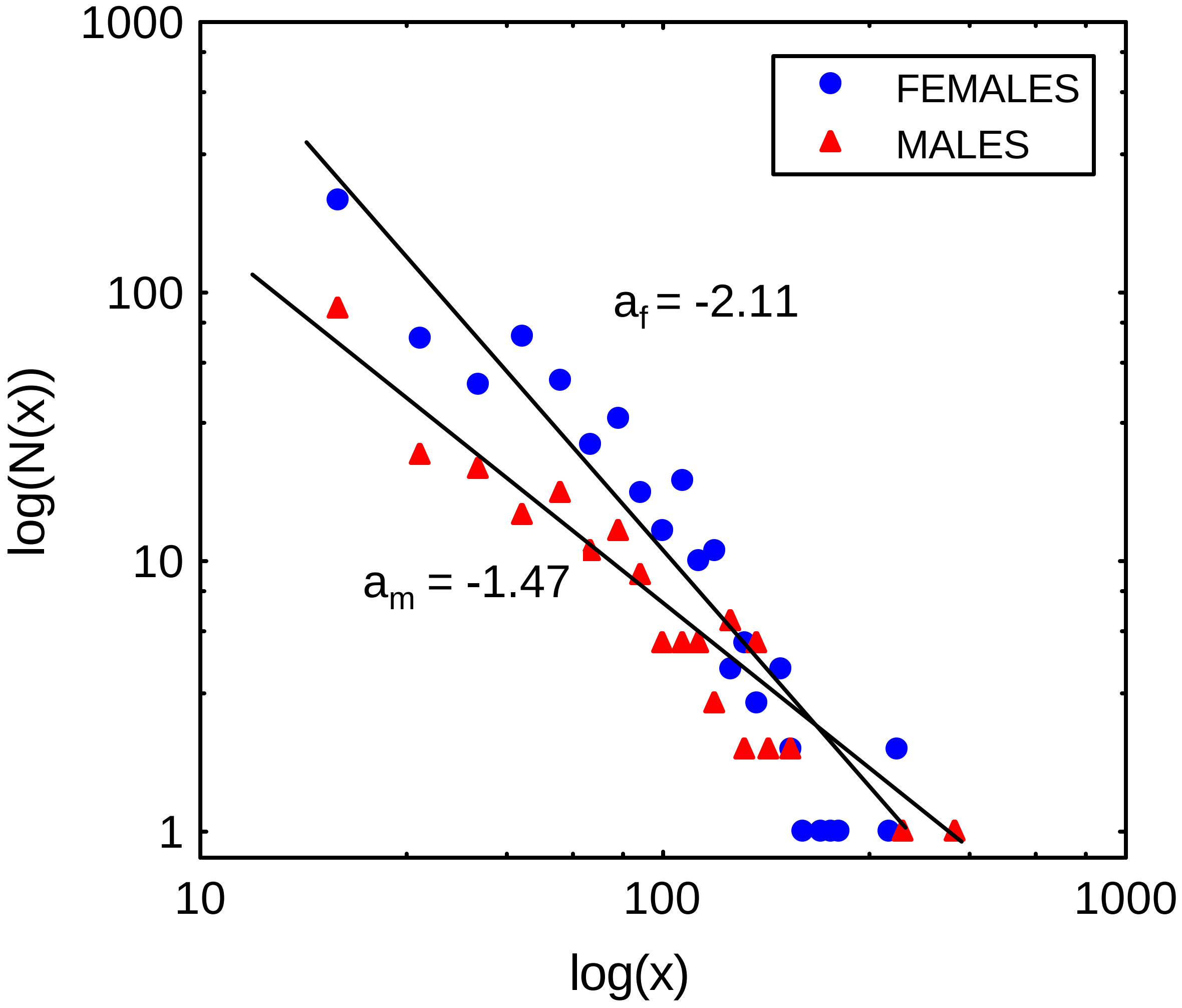